\documentstyle[preprint,aps,tighten]{revtex}

\begin{document}
\title{Heavy Meson Description with a Screened Potential}
\author{P. Gonz\'alez$^{(1)}$, A. Valcarce$^{(2)}$, H.Garcilazo$^{(3)}$, and J.
Vijande$^{(2)}$}
\address{$(1)$ Dpto. de F\' \i sica Te\'orica and IFIC\\
Universidad de Valencia - CSIC, E-46100 Burjassot, Valencia, Spain}
\address{$(2)$ Grupo de F\' \i sica Nuclear \\
Universidad de Salamanca, E-37008 Salamanca, Spain}
\address{$(3)$ Escuela Superior de F\' \i sica y Matem\'aticas \\
Instituto Polit\'ecnico Nacional, Edificio 9,\\
07738 M\'exico D.F., Mexico}
\maketitle

\begin{abstract}
We perform a quark model calculation of the $b\overline{b}$ and $c\overline{c}$
spectra from a screened funnel potential form suggested by unquenched
lattice calculations. A connection between the lattice screening parameter
and an effective gluon mass directly derived from QCD is established.
Spin-spin energy splittings, leptonic widths and radiative decays are also
examined providing a test for the description of the states.

\vspace*{2cm} \noindent Keywords: phenomenological quark models, meson
properties \newline
\newline
\noindent Pacs: 12.39.-x, 14.40.-n, 13.40.Hq
\end{abstract}

\newpage

\section{Introduction}

Heavy quarkonia, $b\overline{b}$ and to a lesser extent 
$c\overline{c}$ (for $t\overline{t}$ the weak decay rates are too
large for these resonances to be observed), are ideal systems to test quark
potential models derived from QCD. Actually for the low-lying states (the
ground state for $c\overline{c}$ and the ground and first excited
states for $b\overline{b}$) one can test the theory itself at the
perturbative regime \cite{YND}. This may provide a quantitative estimate of
the approximations involved when using a potential model interaction derived
from QCD to describe heavy quarkonia.

At very short distances a nonrelativistic coulombic potential, with strength
proportional to $\alpha _{s}$, the quark-gluon-quark coupling constant, is
derived perturbatively from the one-gluon exchange interaction in QCD.
Relativistic corrections involve spin-independent, spin-spin, spin-orbit and
tensor terms. Radiative corrections have been estimated as well. At long
distances nonperturbative effects involving multigluon interactions, in
particular those related to confinement, have to be considered. Lattice
calculations in the so-called quenched approximation (only valence quarks)
derive a heavy quark long range potential linearly dependent on the
interquark distance \cite{BALI}.

The resulting interaction can then be parametrized by a funnel shape
coulomb+linear potential. Such a potential has been applied to get a
good description of the spectrum of spin-triplet states for $c\overline{c}$
and for the first excited states of $b\overline{b}$ \cite{EICH}.
When completed with some relativistic plus radiative corrections it can give
account, through a phenomenological fit of the mass parameters, for the
whole meson spectrum ranging from heavy to light mesons, as well as for the
electromagnetic, strong and weak decays \cite{ISGUR,BHA}.

Further QCD corrections to the coulomb+linear form can be incorporated in
the potential through unquenched (valence+sea quarks) lattice calculations 
\cite{BALI}. The spontaneous creation of light quark pairs may give
rise to a breakup of the color flux tube between the two heavy quarks. It
has been proposed that this translates into a screened funnel potential \cite
{BALI,BORN,DEFG}, say the potential does 
not rise continuously with the interquark
distance but it saturates at the splitting energy of the heavy quark pair. 
Although string breaking can not be said to be definitely confirmed
through lattice calculations \cite{C} a quite rapid cross-over from a linear
rising to a flat potential is well established in SU(2) Yang-Mills theories 
\cite{DEFG}. The analysis of the consequences of such a screening in heavy
quarkonia is our main goal in this article. To this purpose we discuss in
Sec. II the screened form of the potential and connect it to an
effective running gluon mass in QCD. In this form we can obtain from the
underlying theory the value of the screening parameter for a given energy
scale. Then, via a sensible choice of this scale, we calculate in Secs. III
to V the spectra, spin-spin splittings, leptonic widths and radiative decays
of $b\overline{b}$ and $c\overline{c}$ mesons. The dependence of our results
on the chosen scale is studied in Sec. VI. Finally in Sec. VII we summarize
our results.

\section{The screened funnel potential}

The long range behavior of the heavy quark potential can be assigned to the
saturation property of the running QCD coupling constant, $\alpha
_{s}(Q^{2}) $, at the low momentum scale. In a particular framework it
has been shown \cite{PAPA}, through a solution of the Schwinger-Dyson
equation, that when decreasing $Q^{2}$, $\alpha _{s}$ does not increase
continuously but instead it saturates at a constant value $\alpha
_{s}(0)\simeq 0.8$. Explicitly: 
\begin{equation}
\alpha _{s}(Q^{2})=\frac{4\pi }{\beta _{0}\ln \left[
(Q^{2}+4M_{g}^{2}(Q^{2}))/\Lambda ^{2}\right] }  \label{RC}
\end{equation}
where $\beta _{0}=(33-2n_{f})/3$, being $n_{f}$ the number of flavors with
mass much smaller than $Q$. $\Lambda $ is the QCD scale parameter for which
we shall take $300$ MeV hereforth and $M_{g}$ is an effective gluon mass
given by: 
\begin{equation}
M_{g}^{2}(Q^{2})=m_{g}^{2}\,\,\left( \frac{\ln \left[ (Q^{2}+4m_{g}^{2})/%
\Lambda ^{2}\right] }{\ln \left( 4m_{g}^{2}/\Lambda ^{2}\right) 
}\right)^{-12/11} \label{GM}
\end{equation}
with $m_{g}\sim 300$ MeV. Hence the effective gluon mass, $M_{g}$, runs from 
$0$ at $Q^{2}\rightarrow \infty $ to $m_{g}$ at $Q^{2}\rightarrow 0$.
Correspondingly $\alpha _{s}$ runs from $0$ at $Q^{2}\rightarrow \infty$
(asymptotic freedom) to $\sim 0.8$ at $Q^{2}\rightarrow 0$. By implementing
the one-gluon-exchange diagram with $\alpha _{s}(Q^{2})$ as given above but
substituting $M_{g}^{2}(Q^{2})$ by $M_{g}^{2}(Q^{2}\rightarrow 0)\simeq
\Lambda ^{2}$, a long range ($r>>\Lambda ^{-1}$) linear potential comes out 
\cite{YND} (one should keep in mind that fine structure
splittings might require a scalar\cite{SC78} or a mixed scalar-vector
structure for the confinement \cite{KA97}).

This linear tendency gets modified by $q\overline{q}$ pair creation in
between the heavy quarks since a screening of the quark color charge at
large distances takes place. Moreover the effective coulomb coupling is also
affected by the presence of sea quarks. These effects were parametrized in
exploratory lattice studies by a screened funnel 
potential \cite{BORN}. For the sake of simplicity we shall follow
this parametrization though it does not reproduce the rapid
turnover around 1 fm from linearly rising to flat potential
suggested by modern lattice results \cite{DEFG}. For a meson
the potential has the form: 
\begin{equation}
\overline{V}(r)=\left( \overline{\sigma }r-\frac{4}{3}\frac{\overline{\alpha 
}_{s}}{r}\right) \left( \frac{1-e^{-\mu r}}{\mu r}\right) \,  \label{SP}
\end{equation}
where $\mu $ is the screening parameter (with units of an inverse length)
and the hat over the parameters in the potential distinguishes them from the
nonscreened case: $V(r)=\sigma r-\frac{4}{3}\frac{\alpha _{s}}{r}$. $%
\overline{V}(r)$ behaves like a coulomb potential for $r\rightarrow 0$
whereas it tends to $\overline{\sigma }/\mu $ for $r\rightarrow \infty $.

Regarding the value of $\mu $ let us realize that for $\mu \rightarrow 0$
one formally recovers the nonscreened potential with the long distance
linear behavior one had when doing $M_{g}(Q^{2})\sim \Lambda $. On the other
hand the confining part of the potential can be written in the form 
\begin{equation}
\overline{V}_{conf}(r)=\frac{\overline{\sigma }}{\mu }-\overline{\sigma }r%
\frac{e^{-\mu r}}{\mu r}=\overline{V}_{conf}(r\rightarrow \infty )-\overline{%
\sigma }r\frac{e^{-\mu r}}{\mu r}\,  \label{CSP}
\end{equation}
what suggests that $\mu $ should be connected to an exchanged (Yukawa type)
mass. From these considerations we propose this mass to be the effective
gluon mass through the identification 
\begin{equation}
\mu =\Lambda -M_{g}\,  \label{MU}
\end{equation}
so that for a given scale specified by $Q^{2}$ 
\begin{equation}
\overline{V}_{conf}(r)=\frac{\overline{\sigma }}{\mu }-\overline{\sigma }r%
\frac{e^{-\mu r}}{\mu r}=\frac{\overline{\sigma }}{\mu }-\overline{\sigma }r%
\left[ \frac{e^{-(\Lambda -M_{g}(Q^{2}))r}}{(\Lambda -M_{g}(Q^{2}))r}\right]
\,  \label{VCONF}
\end{equation}

This identification establishes a deep connection between the saturation of
the coupling constant and the interquark pair creation mechanism both
effects governed by $M_{g}(Q^{2})$. Therefore $\mu $ runs with $Q^{2}$ so
that $0\simeq \mu (Q^{2}=0)\leq \mu (Q^{2})\leq \mu (Q^{2}\rightarrow \infty
)\simeq$ 1.52 fm$^{-1}$.

The splitting energy of quark and antiquark in the meson is given by $%
\overline{\sigma }/\mu $, the maximum value of the potential. No bound state
can be found for higher energies due to the breaking of the color flux tube
and the most favored subsequent decay in multimeson states. We shall assume
that this splitting energy does not depend on flavor or, more precisely,
that it does not depend on the scale $Q^{2}$. Hence $\overline{\sigma }$
should also run with the scale $Q^{2}$ in such a way that $\overline{\sigma }%
(Q^{2})/\mu (Q^{2})\simeq $ constant. For us $\overline{\sigma}$
will be a free parameter to be determined phenomenologically at a chosen
scale. As a value for comparison we could think of the string tension value
for the nonscreened case, $\sigma \simeq 1000$ MeV fm$^{-1}$ 
\cite{BALI2}. A more physical comparison can be done by evaluating,
from $\overline{\sigma}$, $\mu$, and $\overline{\alpha}_{s}$,
a hadronic scale $R_{0}$ defined through the force between
static quarks at intermediate distances as $F(R_{0}) \, R_{0}^{2}=1.65$.
For the nonscreened case $R_{0}\simeq 0.5$ fm \cite{SOM93}. Concerning
the other parameter of the model, $\overline{\alpha }_{s}$, since we have
established a link between saturation and screening we shall assume 
$\overline{\alpha }_{s}\simeq \alpha _{s}$ as given by Eq. (\ref{RC}) at the
chosen scale.

Our meson hamiltonian will then be written, up to an additive constant, as: 
\begin{equation}
H=m_{q}+m\overline{_{q}}+\frac{p_{q}^{2}}{2m_{q}}+\frac{p_{\overline{q}}^{2}%
}{2m_{\overline{q}}}+\overline{V}(r)\,
\end{equation}
where the constituent quark and antiquark masses, $m_{q}$ and $m\overline{%
_{q}}$, are taken as free parameters.

\section{The \lowercase{$b\overline{b}$, $c\overline{c}$} and 
\lowercase{$c\overline{b}$} spectra.}

In order to apply the screened potential to obtain the $b\overline{b}$ and $c%
\overline{c}$ spectra we should first identify the $Q^{2}$ scale in each
case. For the sake of simplicity and universality let us investigate the
possibility to describe $b\overline{b}$ and $c\overline{c}$ mesons with the
same set of potential parameter values, say by using the same $Q^{2}$ scale.
A sensible guess may be to choose $Q^{2}$ as some intermediate value between
the square quark masses $m_{b}^{2}$ and $m_{c}^{2}$. But it turns out that
our quark model masses are parameters to be determined from the spectra.
Nonetheless we can consider as a reference the central quark mass values 
($\overline{\rm{MS}}$ scheme) quoted by the Particle Data 
Book \cite{PDG}, $m_{b}=4.2$ GeV and 
$m_{c}=1.2$ GeV, and select $Q^{2}=(2.7$ GeV)$^{2}$.

Once $Q^{2}$ has been chosen $\alpha _{s}$ and $\mu $ are determined by Eqs.
(\ref{RC}),(\ref{GM}) and (\ref{MU}). Then we should establish the protocol
to fix the other parameters. We shall fix the quark masses and $\overline{%
\sigma }$ to get the experimental mass for the ground states of $b\overline{b%
}$ and $c\overline{c}$ and the first excited state of $b\overline{b}$.
Indeed this corresponds to a possible definition of the mass for the
confined quarks. Let us note that we do not expect to get the reference 
$\overline{\rm{MS}}$
quark masses since they do not correspond to the same definition but quite
different values will not be compatible with\ the self-consistency of our
scheme. The values of the parameters used are compiled in Table \ref{t1}.
The resulting spectra, square mean root radii and square quark velocities
appear in Tables \ref{t2}, \ref{t3} and \ref{t4}.

Let us note that our hamiltonian is spin independent what means that $S=0$
and $S=1$ states are degenerated. Assuming that the energy difference
between spin singlet and triplet states comes out mainly due to the
spin-spin interaction and taking into account the magnitude of these
splittings and the fact that the matrix element for the spin-spin operator $%
\vec{S}_{i} \cdot \vec{S}_{j}$ is three times bigger for singlets than for
triplets, we shall consider our quark model states as close description to
spin triplets.

The good agreement of the mass prediction with data for $b\overline{b}$ and $%
c\overline{c}$ up to $1$ GeV excitation energy allows an unambiguous
identification of experimental $J^{P}=1^{-}$ states: $1s$, $2s$, $1d$, $3s$
and $4s$ for $b\overline{b\text{,}}$ and $1s$ and $2s$ for $c\overline{c}$.
This is reinforced by the calculation of leptonic width ratios, $\left[
\Gamma _{e^{+}e^{-}}/\Gamma _{e^{+}e^{-}}(1s)\right] $, (see Sec. IV). The
same analysis suggests that for $c\overline{c}$ our $1d$ and $3s$ states
must have some mixing in order to explain the experimental values of the
leptonic width ratios, bigger for $1d$ and smaller for $3s$ than our
approximated results.

For $p$ states and higher energies the situation is much less clear and the
interpretation of our results requires some additional considerations.
First, relativistic effects, specially for the $c\overline{c}$ system (see
the $\left\langle v^{2}/c^{2}\right\rangle $ column in Tables \ref{t2}, \ref
{t3} and \ref{t4}) may be quite relevant: higher order kinetic energy terms
increase in importance when increasing the excitation energy, the spin-orbit
force can be responsible for the $p$ splittings and for differences between $%
L=0$ and $L=2$ states, the tensor interaction can induce $s-d$ mixing, the
color magnetic potential breaks the degeneracy with $0^{-}$ states. Second,
coupling to $D$ and $B$ mesons may become relevant \cite{Upsi}. Third,
there can be some bias in the results associated to the $Q^{2}$ scale
assumed. Anyhow a glance at the experimental leptonic width ratios and their
comparison to our calculation, Tables \ref{t5}, suggests that $\Upsilon
(10860)$ and $\Upsilon (11020)$ should be mostly $s$ states since for $d$
states the predicted ratios would be three orders of magnitude smaller. For $%
\psi (4160)$ and $\psi (4415)$ an $s$ state association seems also to be
favored, Tables \ref{t6} and \ref{t9}.

Concerning $c\overline{b}$ the commented results and the good prediction for
the ground state gives us also confidence about the predicted masses of its
lower excitations.

The quality of the spectra we get is quite similar to that of other quark
model calculations that use strict (nonscreened) confinement \cite
{EICH,ISGUR}. The major difference concerns the possible number of
quark-antiquark bound states which is finite in our case and infinite for
the strict confinement models. This also affects the pattern of energy
differences but mostly for higher excited states. Unfortunately the
uncertainties associated to the calculation of these states are such that a
comparison to data cannot discriminate against any model. The upper energy
cut for the spectrum is given in our model by $m_{q}+m_{\overline{q}}+%
\overline{\sigma }/\mu $. From the values of the parameters we obtain a
splitting energy $\overline{\sigma }/\mu \simeq 2070$ MeV and the energy
thresholds $E_{th}(b\overline{b})\simeq 11400$ MeV, $E_{th}(c\overline{b}%
)\simeq 7973$ MeV and $E_{th}(c\overline{c})\simeq 4547$ MeV. Although these
values have to be taken very cautiously (actually they are significantly
lower than the ones obtained from a Regge analysis \cite{GOLD}) since they
depend on the somewhat arbitrary $Q^2$ scale (see Sec. VI) the qualitative
consequence is clear: the mere existence of a threshold means that a
detected state above it would correspond to a wide resonance, a
glueball or an exotic state. Moreover, it could even correspond to a narrow
resonance provided the existence of selection rules preventing its strong
decay.

It is worth to remark that our quark mass values are smaller than the ones
obtained from strict confinement models ($m_{b}\sim 5000$ MeV, $m_{c}\sim
1700$ MeV) may be indicating that relativistic effects associated to quark
pair creation can be taken into account in an effective way, following the
philosophy of quark model calculations, via a bigger nonrelativistic quark
mass. Concerning the value of $\overline{\sigma }$ (1470 MeV fm$^{-1}$) it
is quite big as compared to commonly accepted values for the string tension $%
\sigma$. However the hadronic scale provided by our model 
$R_{0}\simeq 0.43$ fm is much closer to the nonscreened case. We should
also  keep in mind that $\overline{\sigma}$ is an effective
parameter which may give account of some nonconsidered corrections and that 
$\overline{\sigma}$ runs with $Q^{2}$ in such a way that the
lower the $Q^{2}$ the lower the $\overline{\sigma}$ (see Sec. VI).

\subsection{Singlet-triplet states}

As the screened potential we use has no spin-dependence, spin singlet ($S=0$%
) and triplet $(S=1)$ states are degenerated. If we consider the
color-magnetic spin-spin term coming from the one-gluon-exchange diagram in
QCD and screen it in the same manner than the rest of the potential, i.e., 
\begin{equation}
V_{ss}=\frac{4}{3}\alpha _{s}\frac{8\pi }{3m_{q}m_{\overline{q}}}\,\,\vec{S}%
_{i}\cdot \vec{S}_{j}\,\,\delta (\vec{r})\left( \frac{1-e^{-\mu r}}{\mu r}%
\right) \,  \label{SS}
\end{equation}
we realize that when using it in the Schr\"{o}dinger equation the presence
of $\delta (\vec{r})$ gives rise to an unbound low-energy spectrum. This
difficulty can be overcome substituting the $\delta (\vec{r})$ by a
spreading function depending on a characteristic length. However as this
substitution represents also some kind of screening, avoiding the contact
interaction, the inverse screening length, $\mu $, appearing in Eq. (\ref{SS}%
) could not coincide with the value of $\mu $ for the rest of the potential.
Then we would be left with two new parameters, the spreading length and a
modified inverse screening length, to fit one established and two to be
confirmed experimental splitting energies: $M(J/\psi )-M[\eta _{c}(1s)]$, $%
M[\psi (2s)]-M[\eta _{c}(2s)]$ and $M[\Upsilon (1s)]-M[\eta _{b}(1s)]$, what
does not seem to be a big deal. Instead we can try to preserve the
simplicity of our model and apply first order perturbation theory. Thus the
splitting energies obtained from $J/\psi $, $\psi (2s)$ and $\Upsilon (1s)$
are expressed as 
\begin{equation}
\Delta E_{ss}=\frac{4}{3}\alpha _{s}\frac{8\pi }{3m_{q}m_{\overline{q}}}%
\left| \Psi (0)\right| ^{2}\,  \label{ESS}
\end{equation}
where $\Psi (0)$ stands for the wave function at the origin for $J/\psi $, $%
\psi (2s)$ and $\Upsilon (1s)$ respectively. So we predict: 
\begin{equation}
M(J/\psi )-M[\eta _{c}(1s)]\simeq 118.6\,\,{\rm MeV}\,
\end{equation}
\begin{equation}
M[\Upsilon (1s)]-M[\eta _{b}(1s)]\simeq 104.3\,\,{\rm MeV}
\end{equation}
\begin{equation}
M[\psi (2s)]-M[\eta _{c}(2s)]\simeq 69.4\,\,{\rm MeV}\,
\end{equation}
to be compared to the experimental data\cite{PDG}: 
\begin{equation}
M(J/\psi )-M[\eta _{c}(1s)]=117.9\pm 2.1\text{ \ }{\rm MeV}\,
\end{equation}
\begin{equation}
M[\Upsilon (1s)]-M[\eta _{b}(1s)]=160\pm 40\text{ \ }{\rm MeV}\,
\end{equation}
and \cite{BELLE}
\begin{equation}
M[\psi (2s)]-M[\eta _{c}(2s)]=64\pm 12\text{ \ }{\rm MeV} \, .
\end{equation}
Though we get a remarkable agreement, as the magnitudes of the
splittings are not much smaller than the nonperturbed Schr\"{o}dinger
equation eigenvalues we have [620 MeV for $J/\psi $, 1202 MeV for $\psi (2s)$
and 131 MeV for $\Upsilon (1s)$], specially for $b\overline{b}$, the first
order perturbed values should not be taken for granted. Anyhow we can
tentatively make some additional predictions for the next two splittings: 
\begin{equation}
M[\Upsilon (2s)]-M[\eta _{b}(2s)]\simeq 50\,\,{\rm MeV}\,
\end{equation}
\begin{equation}
M[\psi (3s)]-M[\eta _{c}(3s)]\simeq 52\,\,{\rm MeV}\,
\end{equation}

\section{Leptonic widths}

A complete calculation of $V$ (vector meson)$\rightarrow e^{+}e^{-}$ widths
involves radiative and relativistic contributions out of the scope of our
quark model estimation. Fortunately these corrections can be factorized \cite
{BTYE}. Thus for $s$ states we can write the leptonic width $\Gamma
_{e^{+}e^{-}}$ as 
\begin{equation}
\Gamma _{e^{+}e^{-}}(ns)=\Gamma _{e^{+}e^{-}}^{(0)}(ns)\left[ 1-\frac{%
16\alpha _{s}}{3\pi }+\Delta (ns)\right] \,  \label{LW}
\end{equation}
The second $(-16\alpha _{s}/3\pi )$ and third $[(\Delta (ns)]$ terms in the
parenthesis on the right hand side stand for the leading order radiative and
higher order radiative+relativistic corrections respectively and 
\begin{equation}
\Gamma _{e^{+}e^{-}}^{(0)}(ns)=\frac{16\pi e_{q}^{2}\alpha ^{2}}{M_{ns}^{2}}%
\left| \Psi _{ns}(0)\right| ^{2}\,  \label{LW0}
\end{equation}
where $e_{q}$ is the electric quark charge, $\alpha $ the fine structure
constant, $M_{ns}$ the mass of the $ns$ state and $\Psi _{ns}(0)$ its wave
function at the origin.

From Eqs. (\ref{ESS}), (\ref{LW}), and (\ref{LW0}) we can establish an
approximate relation between the correction factors and the quark masses
through experimental quantities: 
\begin{equation}
\frac{m_{c}^{2}}{m_{b}^{2}}\frac{\left[ 1-\frac{16\alpha _{s}}{3\pi }+\Delta
(1s)\right] _{c\overline{c}}}{\left[ 1-\frac{16\alpha _{s}}{3\pi }+\Delta
(1s)\right] _{b\overline{b}}}\simeq \frac{\left[ \Gamma _{e^{+}e^{-}}(1s)%
\right] _{c\overline{c}}}{\left[ \Gamma _{e^{+}e^{-}}(1s)\right] _{b%
\overline{b}}}\frac{\left[ M(\Upsilon (1s))-M(\eta _{b}(1s))\right] }{\left[
M(J/\psi )-M(\eta _{c}(1s))\right] }\left[ \frac{M(J/\psi )}{M(\Upsilon (1s))%
}\right] ^{2}\,
\end{equation}

From Eqs. (\ref{LW}) and (\ref{LW0}) it is also clear that a comparison of
the calculated $\Gamma _{e^{+}e^{-}}^{(0)}(ns)$ to data will give us the
effective magnitude of $\Delta $ needed in our model. So for $b\overline{b}$
we obtain $\Gamma _{e^{+}e^{-}}^{(0)}(1s)=2.13$ keV and $\Delta (1s)=0.16$
and for $c\overline{c}$ we have $\Gamma _{e^{+}e^{-}}^{(0)}(1s)=6.37$ keV
and $\Delta (1s)=0.36$. We observe that higher order corrections are much
more important for the $c\overline{c}$ system as expected.

As $\Delta $ includes relativistic corrections it depends on the $ns$ state.
Inasmuch this dependence is not very strong we could assume a mean value for
all these states. Then we could consider the ratio: 
\begin{equation}
\frac{\Gamma _{e^{+}e^{-}}(ns)}{\Gamma _{e^{+}e^{-}}(1s)}\simeq \frac{\Gamma
_{e^{+}e^{-}}^{(0)}(ns)}{\Gamma _{e^{+}e^{-}}^{(0)}(1s)}=\frac{\left| \Psi
_{ns}(0)\right| ^{2}}{\left| \Psi _{1s}(0)\right| ^{2}}\frac{M_{1s}^{2}}{%
M_{ns}^{2}}\,  \label{RATIO}
\end{equation}
where correction factors cancel. Indeed as far as the masses of the states
agree with data this is a test of the ratio of the wave functions at the
origin.

Moreover from Eqs. (\ref{ESS}) and (\ref{RATIO}) we can predict 
\begin{equation}
\left[ \frac{\Gamma _{e^{+}e^{-}}(2s)}{\Gamma _{e^{+}e^{-}}(1s)}\right] _{c%
\overline{c}}\simeq \left[ \frac{M(\psi (2s))-M(\eta _{c}(2s))}{M(J/\psi
)-M(\eta _{c}(1s))}\right] \left[ \frac{M(J/\psi )}{M(\psi (2s))}\right]
^{2}\,
\end{equation}
close to the experimental data, 
\begin{eqnarray}
&&\left[ \frac{\Gamma _{e^{+}e^{-}}(2s)}{\Gamma _{e^{+}e^{-}}(1s)}\right] _{c%
\overline{c}}=0.41\pm 0.07\,, \\
&&\left[ \frac{M(\psi (2s))-M(\eta _{c}(2s))}{M(J/\psi )-M(\eta _{c}(1s))}%
\right] \left[ \frac{M(J/\psi )}{M(\psi (2s))}\right] ^{2}=0.55\pm 0.04\,.
\end{eqnarray}
For $d$ states we use: 
\begin{equation}
\Gamma _{e^{+}e^{-}}^{(0)}(nd)=\frac{25e_{q}^{2}\alpha ^{2}}{%
2m_{q}^{4}M_{nd}^{2}}\left| R_{nd}^{"}(0)\right| ^{2}\,
\end{equation}
where $R_{nd}^{"}(0)$ stands for the second derivative of the radial wave
function at the origin.

Our results, shown in Tables \ref{t5} and \ref{t6}, are quite encouraging,
they seem to confirm the correctness of our assumptions about the validity
of the prediction of ratios and support the state identification done in
Sec. III (let us comment that there could be some bias due to the $Q^{2}$
scale value chosen, see Tables \ref{t8} and \ref{t9}, Sec. VI).

\section{E1 and M1 decay widths}

According to the identification of our states with spin triplets done in
Sec. III, we shall center in $E1$ decays in the $b\overline{b}$ and $c%
\overline{c}$ sectors involving only triplet states, for instance $\chi
_{c_{J}}(1p)\rightarrow \gamma J/\psi $, $\Upsilon (2s)\rightarrow \gamma
\chi _{b_{J}}(1p)$, .... Then for $i\rightarrow \gamma f$, [$i$($f$) stands
for the initial(final) meson], using a single quark current operator and the
nonrelativistic approximation, we can write for the width 
\begin{equation}
\Gamma _{if}^{E1}=\frac{4}{27}e_{q}^{2}\alpha
k_{if}^{3}(2J_{f}+1)D_{if}^{2}\,
\end{equation}
where $k_{if}$ is the photon energy or momentum, $J_{f}$ is the total
angular momentum of the final meson and $D_{if}$ the transition matrix
element 
\begin{equation}
D_{if}=\int_{0}^{\infty }dru_{i}(r)\frac{3}{k_{if}}\left[ \frac{k_{if}r}{2}%
j_{0}\left( \frac{k_{if}r}{2}\right) -j_{1}\left( \frac{k_{if}r}{2}\right) %
\right] u_{f}(r)\,
\end{equation}
being $u_{i,f}(r)$ the reduced radial wave functions for the initial and
final mesons respectively and $j_{0}$ and $j_{1}$ spherical Bessel
functions. In the limit $k_{if}r\rightarrow 0$ one recovers the dipole form: 
\begin{equation}
D_{if}\rightarrow \int_{0}^{\infty }dru_{i}(r)ru_{f}(r)\,
\end{equation}

In our simple quark model all the $J$ states for $p$ waves, $\chi _{J}$, are
degenerated. Let us assume that this degeneration can be dropped out by the
consideration of some perturbative interaction such as a spin-orbit one and
that first order perturbation theory can give us the right masses. This is
quite reasonable since the needed splittings are about one tenth of the
energy eigenvalues. Hence we shall use our wave functions (first order
perturbation theory does not affect the wave functions) altogether with the
experimental masses to get the $E1$ rates.

Certainly there are relativistic and radiative corrections to the formulas
above (involving wave function corrections also). We shall take as a
criterium to estimate its importance the value of $k_{if}R$ ($R$ can be
assimilated to the mean radius of the initial meson) as compared to $1$.
Then for the transitions we consider next and from Tables \ref{t2} and \ref
{t3} we have $\left( k_{if}R\right) _{b\overline{b}}\simeq 0.35$ and $\left(
k_{if}R\right) _{c\overline{c}}\geq 1.$ Hence we expect a much more
reasonable description for $b\overline{b}$ than for $c\overline{c}$. This
expectation is confirmed by our results as can be checked in Table \ref{t7}:
the $b\overline{b}$ data are quite well described whereas the $c\overline{c}$
widths deviate by a factor of about $2$.

The much bigger quality of the results we obtain for $b\overline{b}$ as
compared to the ones obtained with a simple nonscreened linear potential 
\cite{EICH} points out the importance of screening effects. On the other
hand by taking into account that with a more refined model without screening 
\cite{ISGUR} both $b\overline{b}$ and $c\overline{c}$ widths are correctly
described we can interpret that the effect of screening is taken into
account in this model in an effective manner through the fitted parameters.

Less satisfactory from the experimental point of view is the situation for $%
M1$ decays since there are only a few experimental $c\overline{c}$ data
available [$J/\psi \rightarrow \gamma \eta _{c}(1s)$ and $\psi
(2s)\rightarrow \gamma \eta _{c}(1s)$]. The nonrelativistic theoretical
expression for the width is: 
\begin{equation}
\Gamma _{if}^{M1}=\frac{16}{3}\left( \frac{e_{q}}{2m_{q}}\right) ^{2}\mu
_{q}^{2}\alpha k_{if}^{3}(2J_{f}+1)M_{if}^{2}\,
\end{equation}
with the transition matrix element $M_{if}$ given by 
\begin{equation}
M_{if}=\int_{0}^{\infty }dru_{i}(r)j_{0}\left( \frac{k_{if}r}{2}\right)
u_{f}(r)\,
\end{equation}
and where we have used an effective magnetic factor $\mu _{q}^{2}$ to take
into account corrections to the quark magnetic moment beyond the Dirac
particle term.

As the $J/\psi -\eta _{c}(1s)$ and the $\psi (2s)-\eta _{c}(2s)$ mass
splittings are reasonably well reproduced by first order perturbation theory
we follow the same philosophy as above and assume the same wave function for 
$J/\psi $ and $\eta _{c}(1s)$ and analogously for $\psi (2s)$ and $\eta
_{c}(2s)$. Let us realize that if we fit $\mu _{q}^{2}$ from data this
factor can effectively incorporate corrections of the same type mentioned
for the $E1$ case as well as corrections that would come from a better wave
function description (specially for singlet states).

Then, for $J/\psi -\eta _{c}(1s)$, by using that $M_{if}\simeq 1$ we can
estimate $\Gamma _{J/\psi \rightarrow \gamma \eta _{c}(1s)}\simeq
4.26\,\,\mu _{c}^{2}$. By comparing with the experimental value, $\Gamma
_{J/\psi \rightarrow \gamma \eta _{c}(1s)}=1.2\pm 0.4$ keV we get $\mu
_{c}\simeq 0.52$. We can compare this value with the corresponding to the
second order correction to the magnetic moment operator \cite{RISKA} $\left(
1-2v_{c}^{2}/3c^{2}\right) \simeq 0.8$. Since our $\mu _{c}$ has a very
effective character the value determined from $\psi (2s)\rightarrow \gamma
\eta _{c}(1s)$ could be different. If keeping in mind these caveats we
assume the same $\mu _{c}^{2}$ we can predict $\Gamma _{\psi (2s)\rightarrow
\gamma \eta _{c}(1s)}\simeq 1.24$ keV to be compared to $\Gamma _{\psi
(2s)\rightarrow \gamma \eta _{c}(1s)}=0.8\pm 0.3$ keV.

Let us also realize that we can get rid of the presence of $\mu _{q}$ by
taking the ratio of the widths. Thus we can predict for the non-measured yet 
$\psi (2s)\rightarrow \eta _{c}(2s)$ transition: 
\begin{equation}
\frac{\Gamma _{\psi (2s)\rightarrow \gamma \eta _{c}(2s)}}{\Gamma _{J/\psi
\rightarrow \gamma \eta _{c}(1s)}}\simeq \left( \frac{k_{\psi
(2s)\rightarrow \gamma \eta _{c}(2s)}}{k_{J/\psi \rightarrow \gamma \eta
_{c}(1s)}}\right) ^{3}\simeq 0.5
\end{equation}
since $M_{if}\simeq 1$ for both transitions.

Finally for $b\overline{b}$, since we expect relativistic corrections to be
much less important, by assuming $\mu _{b}\simeq \left(
1-2v_{b}^{2}/3c^{2}\right) =0.94$ (very close to $1$, the value for a Dirac
particle), we predict $\Gamma _{\Upsilon (1s)\rightarrow \gamma \eta
_{b}(1s)}=0.17$ keV (we have used $M[\eta _{b}(1s)]=9300$ MeV \cite{PDG};
by including the experimental range $M[\eta _{b}(1s)]=9300\pm 20\pm
20 $ MeV we have $\Gamma _{\Upsilon (1s)\rightarrow \gamma \eta
_{b}(1s)}=0.07-0.33$ keV).

\section{The $Q^{2}$ scale}

Although our results are good enough to provide an {\it ad hoc}
justification for the universal $Q^{2}$ chosen for $b\overline{b}$ and $c%
\overline{c}$ there can be convenient to try to justify it from more
physical grounds. If we write $Q^{2}=-q^{2}$, being $q$ is the momentum
transfer between quark and antiquark, we have, in the center of mass system, 
$Q^{2}=4\vec{p}_{q}^{\,\,2}$ where $\vec{p}_{q}$ is the quark trimomentum $%
\vec{p}_{q}^{\,\,2}=m_{q}^{2}v_{q}^{2}/\left( 1-v_{q}^{2}\right) $. By
taking as a reference the values of quark masses and velocities previously
obtained for $1s$ states we get $Q_{b\overline{b}}^{2}\simeq \left( 2.9\,\,%
{\rm GeV}\right) ^{2}$ and $Q_{c\overline{c}}^{2}\simeq \left( 2.4\,{\rm GeV}%
\right) ^{2}$, pretty close to the $Q^{2}$ scale value we guessed $%
Q^{2}=(2.7 $ GeV)$^{2}$. Let us also note that if we had used the
nonrelativistic expressions for the trimomentum we would have got $Q_{b%
\overline{b}}^{2}\simeq \left( 2.8\,\,{\rm GeV}\right) ^{2}$ and $Q_{c%
\overline{c}}^{2}\simeq \left( 1.3\,\,{\rm GeV}\right) ^{2}$ what reflects
the much more relativistic character of $c\overline{c}$. Anyhow as we are
using a nonrelativistic scheme it can be of help to analyze the variation of
our results when a $Q^{2}\simeq \left( 1.3\,\,{\rm GeV}\right) ^{2}$ scale
is chosen for $c\overline{c}$. Actually it is possible to choose $Q_{c%
\overline{c}}^{2}=m_{c}^{2}=\left( 1357\,\,{\rm MeV}\right) ^{2}$, so that $%
\mu =0.556$ fm$^{-1}$ and $\overline{\sigma }=1175$ MeV fm$^{-1}$,
very close to the string tension value. Furthermore $\overline{\sigma}%
/\mu =2113$ MeV, very close to the value obtained with the former universal
scale what seems to confirm our assumption about its constancy. The results
are shown in Tables \ref{t8} and \ref{t9}. Due to the bigger value of the
quark mass the upper energy threshold increases with respect to our former
value (4827 MeV $vs$ 4547 MeV). From the comparison with Tables \ref{t2}, 
\ref{t3}, \ref{t4}, \ref{t5} and \ref{t6}, it is clear that the effect of
lowering the scale translates in a smaller size and a bigger wave function
value (for $ns$ waves) at the origin. The rates $\Gamma
_{e^{+}e^{-}}(ns)/\Gamma _{e^{+}e^{-}}(1s)$ get reduced approaching data. $%
E1 $ decay rates change also, although by a small amount, in the direction
pointed by data. These tendencies though not very significant seem to favor
for $c\overline{c}$ a lower scale than the one used in Sec. III.

\section{Summary}

We have studied heavy quarkonia from a simple screened funnel quark
potential model. The form of the potential is suggested by unquenched
lattice calculations. The screening inverse length, i. e. the parameter of
the potential controlling the onset of screening effects, has been related
to an effective gluon mass derived from QCD. As such a mass is also
responsible for the saturation property of the running coupling constant in
QCD an underlying connection between saturation and screening comes out. The
model provides a quite accurate description of $b\overline{b}$
masses, spin-spin splittings, leptonic widths and radiative decays. This may
be pointing out the relevance of string breaking effects although the more
genuine predictions associated to screening, say the finiteness of the
quark-antiquark bound state spectrum, its corresponding upper energy
threshold and the breaking color flux tube energy between quarks, remain to
be tested in the future since current high energy excitation data do not
allow to extract neither quantitative nor qualitative conclusions about
their validity.

For $c\overline{c}$ the implementation of other relativistic corrections,
apart from screening, seems to be essential to approach data as confirmed by
the accurate description provided for more refined models. There could be,
in our opinion, interesting to implement screening effects in such kind of
models to try to disentangle the role played by relativistic effects of
different character.

Let us say finally that the model can be also applied to light hadrons,
mesons and baryons, but at the price of introducing more effectiveness in
the values of the parameters, losing the stringent connection established
with the underlying theory.

\acknowledgments
We are specially grateful to Dr. M.A. Sanchis for providing us with
phenomenological information. This work has been partially funded by
Ministerio de Ciencia y Tecnolog{\'{i}}a under Contract No. BFM2001-3563, by
Junta de Castilla y Le\'{o}n under Contract No. SA-109/01, and by EC-RTN
(Network ESOP) under Contract No. HPRN-CT-2000-00130.

\begin{table}[tbp]
\caption{Parameters of the model.}
\label{t1}
\begin{tabular}{cccc}
& $\overline\alpha_s$ & 0.317 &  \\ 
& $\mu$ ({\rm fm$^{-1}$}) & 0.71 &  \\ 
& $\overline\sigma$ ({\rm MeV fm$^{-1}$}) & 1470 &  \\ 
& $m_{b}$ ({\rm MeV}) & 4664.5 &  \\ 
& $m_{c}$ ({\rm MeV}) & 1238.5 & 
\end{tabular}
\end{table}

\begin{table}[tbh]
\caption{$b\overline{b}$ spin-triplet bound state masses (in MeV) and properties. We
denote by a '*' the states used to fit the parameters and by a '$\dagger $'
experimental states with different possibilities for the orbital angular
momentum, it could be $L=0$ and/or $L=2$. Experimental data are taken from
Ref. \protect\cite{PDG} except for that quoted by a '**' taken from Ref. 
\protect\cite{CLEO}. 
$p$ states are taken as the centroid
of the $np_0$, $np_1$ and $np_2$ experimental data.}
\begin{tabular}{ccccc}
& Mass & Exp. & $\left\langle {v^{2}/c^{2}}\right\rangle $ & $\left\langle
r^{2}\right\rangle ^{1/2}$ (fm) \\ \hline
$1s$ & 9460$^{\,\ast }$ & $9460.30\pm 0.26$ & 0.09 & 0.23 \\ 
$2s$ & 10023$^{\,\ast }$ & $10023.26\pm 0.31$ & 0.08 & 0.51 \\ 
$1d$ & 10159 & $10162.2\pm 1.6^{\,\ast \ast }$ & 0.07 & 0.55 \\ 
$3s$ & 10350 & $10355.2\pm 0.5$ & 0.08 & 0.78 \\ 
$2d$ & 10436 &  & 0.08 & 0.82 \\ 
$4s$ & 10588 & $10580.0\pm 3.5$ & 0.09 & 1.05 \\ 
$3d$ & 10649 &  & 0.08 & 1.10 \\ 
$5s$ & 10773 & $10865\pm 8^{\,\dagger }$ & 0.09 & 1.33 \\ 
$4d$ & 10819 &  & 0.08 & 1.40 \\ 
$6s$ & 10921 & $11019\pm 8^{\,\dagger }$ & 0.08 & 1.60 \\ 
$5d$ & 10956 &  & 0.08 & 1.70 \\ 
$7s$ & 11040 &  & 0.08 & 1.90 \\ \hline\hline
$1p$ & 9909 & $9900.1\pm 0.5$ & 0.07 & 0.40 \\ 
$2p$ & 10262 & $10260.0\pm 0.5$ & 0.08 & 0.70 \\ 
\end{tabular}
\label{t2}
\end{table}

\begin{table}[tbh]
\caption{$c\overline{c}$ spin-triplet bound state masses (in MeV) and properties. We
denote by a '*' the states used to fit the parameters and by a '$\dagger $'
the experimental states with different possibilities for the orbital angular
momentum, it could be $L=0$ and/or $L=2$. Experimental data are taken from
Ref. \protect\cite{PDG}.
$p$ states are taken as the centroid
of the $np_0$, $np_1$ and $np_2$ experimental data.}
\begin{tabular}{ccccc}
& Mass & Exp. & $\left\langle {v^{2}/c^{2}}\right\rangle $ & $\left\langle
r^{2}\right\rangle ^{1/2}$ (fm) \\ \hline
$1s$ & 3097$^{\,\ast }$ & $3096.87\pm 0.04$ & 0.28 & 0.46 \\ 
$2s$ & 3679 & $3685.96\pm 0.09$ & 0.32 & 0.97 \\ 
$1d$ & 3794 & $3769.9\pm 2.5^{\,\dagger }$ & 0.31 & 1.00 \\ 
$3s$ & 4023 & $4040\pm 10^{\,\dagger }$ & 0.31 & 1.50 \\ 
$2d$ & 4093 &  & 0.31 & 1.60 \\ 
$4s$ & 4248 & $4159\pm 20^{\,\dagger }$ & 0.29 & 2.10 \\ 
$3d$ & 4294 &  & 0.27 & 2.20 \\ 
$5s$ & 4397 & $4415\pm 6^{\,\dagger }$ & 0.23 & 2.95 \\ 
$4d$ & 4427 &  & 0.22 & 3.10 \\ 
$6s$ & 4494 &  & 0.21 & 3.70 \\ \hline\hline
$1p$ & 3517 & $3525.3\pm 0.2$ & 0.12 & 0.75 \\ 
$2p$ & 3914 &  & 0.20 & 1.3 \\ 
\end{tabular}
\label{t3}
\end{table}

\begin{table}[tbh]
\caption{$b\overline{c}$ spin-triplet bound state masses (in MeV) and properties.
Experimental data are taken from Ref. \protect\cite{PDG}.}
\begin{tabular}{ccccc}
& Mass & Exp. & $\left\langle {v^{2}/c^{2}}\right\rangle $ & $\left\langle
r^{2}\right\rangle ^{1/2}$ (fm) \\ \hline
$1s$ & 6362 & $6400\pm 400$ & 0.18 & 0.37 \\ 
$2s$ & 6927 &  & 0.20 & 0.79 \\ 
$1d$ & 7045 &  & 0.19 & 0.82 \\ 
$3s$ & 7269 &  & 0.20 & 1.20 \\ 
$2d$ & 7342 &  & 0.20 & 1.25 \\ 
$4s$ & 7506 &  & 0.20 & 1.70 \\ 
$3d$ & 7556 &  & 0.19 & 1.71 \\ 
$5s$ & 7677 &  & 0.17 & 2.16 \\ 
$4d$ & 7713 &  & 0.17 & 2.24 \\ 
$6s$ & 7799 &  & 0.15 & 2.78 \\ 
$5d$ & 7825 &  & 0.15 & 2.89 \\ 
$7s$ & 7884 &  & 0.12 & 3.56 \\ \hline\hline
&  &  &  &  \\ 
$1p$ & 6779 &  & 0.18 & 0.60 \\ 
&  &  &  &  \\ 
&  &  &  &  \\ 
$2p$ & 7163 &  & 0.20 & 1.02 \\ 
&  &  &  & 
\end{tabular}
\label{t4}
\end{table}

\begin{table}[tbh]
\caption{Leptonic widths $\Gamma _{e^{+}e^{-}}$ (in keV) for $b\overline{b}$%
. We denote by a '$\dagger $' the experimental states with different
possibilities for the orbital angular momentum, it could be $L=0$ and/or $%
L=2 $. Experimental data are taken from Ref. \protect\cite{PDG}.}
\label{t5}
\begin{tabular}{ccccc}
& $\Gamma _{e^{+}e^{-}}^{(0)}\left( 1-{\frac{{16\alpha _{s}}}{{3\pi }}}%
\right) $ & $\left( \Gamma _{e^{+}e^{-}}\right) _{exp}$ & $\Gamma
_{e^{+}e^{-}}^{(0)}/\Gamma _{e^{+}e^{-}}^{(0)}(1s)$ & $\left[ \Gamma
_{e^{+}e^{-}}/\Gamma _{e^{+}e^{-}}(1s)\right] _{exp}$ \\ \hline
$1s$ & 0.98 & $1.32\pm 0.05$ & 1 & 1 \\ 
$2s$ & 0.41 & $0.520\pm 0.032$ & 0.42 & $0.41\pm 0.07$ \\ 
$1d$ & $3.7\times 10^{-4}$ &  & $3.8\times 10^{-4}$ &  \\ 
$3s$ & 0.27 & seen & 0.27 & seen \\ 
$2d$ & $5.8\times 10^{-4}$ &  & $5.9\times 10^{-4}$ &  \\ 
$4s$ & 0.20 & $0.248\pm 0.031$ & 0.21 & $0.19\pm 0.03$ \\ 
$3d$ & $6.7\times 10^{-4}$ &  & $6.8\times 10^{-4}$ &  \\ 
$5s$ & 0.16 & $0.31\pm 0.07^{\,\dagger }$ & 0.16 & $0.24\pm 0.06^{\,\dagger
} $ \\ 
$4d$ & $7.9\times 10^{-4}$ &  & $8.1\times 10^{-4}$ &  \\ 
$6s$ & 0.12 & $0.130\pm 0.030^{\,\dagger }$ & 0.13 & $0.10\pm
0.03^{\,\dagger }$ \\ 
$5d$ & $7.1\times 10^{-4}$ &  & $7.3\times 10^{-4}$ &  \\ 
$7s$ & 0.10 &  & 0.10 & 
\end{tabular}
\end{table}

\begin{table}[tbh]
\caption{Leptonic widths $\Gamma _{e^{+}e^{-}}$ (in keV) for $c\overline{c}$%
. We denote by a '$\dagger $' the experimental states with different
possibilities for the orbital angular momentum, it could be $L=0$ and/or $%
L=2 $. Experimental data are taken from Ref. \protect\cite{PDG}.}
\label{t6}
\begin{tabular}{ccccc}
& $\Gamma _{e^{+}e^{-}}^{(0)}\left( 1-{\frac{{16\alpha _{s}}}{{3\pi }}}%
\right) $ & $\left( \Gamma _{e^{+}e^{-}}\right) _{exp}$ & $\Gamma
_{e^{+}e^{-}}^{(0)}/\Gamma _{e^{+}e^{-}}^{(0)}(1s)$ & $\left[ \Gamma
_{e^{+}e^{-}}/\Gamma _{e^{+}e^{-}}(1s)\right] _{exp}$ \\ \hline
$1s$ & 2.94 & $5.3\pm 0.4$ & 1 & 1 \\ 
$2s$ & 1.22 & $2.19\pm 0.15$ & 0.42 & $0.41\pm 0.06$ \\ 
$1d$ & 0.026 & $0.26\pm 0.04^{\,\dagger }$ & $8.8\times 10^{-3}$ & $(5.0\pm
1.1)\times 10^{-2^{\,\dagger }}$ \\ 
$3s$ & 0.76 & $0.75\pm 0.15^{\,\dagger }$ & 0.26 & $0.15\pm 0.04^{\,\dagger
} $ \\ 
$2d$ & 0.03 &  & $1.1\times 10^{-2}$ &  \\ 
$4s$ & 0.43 & $0.77\pm 0.23^{\,\dagger }$ & 0.15 & $0.15\pm 0.05^{\,\dagger
} $ \\ 
$3d$ & 0.03 &  & $9\times 10^{-3}$ &  \\ 
$5s$ & 0.27 & $0.47\pm 0.10^{\,\dagger }$ & 0.09 & $0.09\pm 0.03^{\,\dagger
} $ \\ 
$4d$ & 0.02 &  & $7.9\times 10^{-3}$ & 
\end{tabular}
\end{table}

\begin{table}[tbh]
\caption{E1 decay widths for $b\overline{b}$ and $c\overline{c}$ (in keV).
Experimental data are taken from Ref. \protect\cite{PDG}.}
\label{t7}
\begin{tabular}{cccccccc}
\multicolumn{3}{c}{$b\overline{b}$} & \multicolumn{3}{c}{$c\overline{c}$} & 
&  \\ \hline
Transition & $\Gamma_{E1}$ & $\Gamma_{exp}$ & Transition & $\Gamma_{E1}$ & $%
\Gamma_{exp}$ &  &  \\ \hline
$\Upsilon (2s) \to \gamma \chi_{b_0} (1P)$ & 1.46 & $1.7 \pm 0.5$ & $%
\chi_{c_0} (1P) \to \gamma J / \psi(1s)$ & 63.94 & $169 \pm 51$ &  &  \\ 
$\Upsilon (2s) \to \gamma \chi_{b_1} (1P)$ & 2.27 & $3.0 \pm 0.8$ & $%
\chi_{c_1} (1P) \to \gamma J/\psi(1s)$ & 125.07 & $295 \pm 71$ &  &  \\ 
$\Upsilon (2s) \to \gamma \chi_{b_2} (1P)$ & 2.32 & $3.1 \pm 0.8$ & $%
\chi_{c_2} (1P) \to \gamma J/\psi(1s)$ & 161.64 & $392 \pm 73$ &  &  \\ 
\hline
$\Upsilon (3s) \to \gamma \chi_{b_0} (2P)$ & 1.71 & $1.4 \pm 0.4$ & $\psi
(2s) \to \gamma \chi_{c_0} (1P)$ & 62.56 & $26 \pm 5$ &  &  \\ 
$\Upsilon (3s) \to \gamma \chi_{b_1} (2P)$ & 2.80 & $3.0 \pm 0.6$ & $\psi
(2s) \to \gamma \chi_{c_1} (1P)$ & 59.39 & $26 \pm 4$ &  &  \\ 
$\Upsilon (3s) \to \gamma \chi_{b_2} (2P)$ & 3.04 & $3.0 \pm 0.6$ & $\psi
(2s) \to \gamma \chi_{c_2} (1P)$ & 42.39 & $21 \pm 4$ &  & 
\end{tabular}
\end{table}

\begin{table}[tbh]
\caption{$c\overline{c}$ spin-triplet 
bound state masses (in MeV) and properties from a
screened funnel hamiltonian with parameters: $\overline{\protect\alpha }%
_{s}= $ 0.451, $\protect\mu =$ 0.556 fm$^{-1}$, $\overline{\protect\sigma }=$
1175 MeV fm$^{-1}$, and $m_{c}=$ 1357 MeV. We denote by a '*' the states
used to fit the parameters and by a '$\dagger $' the experimental states
with different possibilities for the orbital angular momentum, it could be $%
L=0$ and/or $L=2$. Experimental data are taken from Ref. \protect\cite{PDG}.
$p$ states are taken as the centroid
of the $np_0$, $np_1$ and $np_2$ experimental data.}
\label{t8}
\begin{tabular}{ccccc}
& Mass & Exp. & $\left\langle {v^{2}/c^{2}}\right\rangle $ & $\left\langle
r^{2}\right\rangle ^{1/2}$ (fm) \\ \hline
$1s$ & 3097$^{\,\ast }$ & $3096.87\pm 0.04$ & 0.28 & 0.42 \\ 
$2s$ & 3686$^{\,\ast }$ & $3685.96\pm 0.09$ & 0.29 & 0.92 \\ 
$1d$ & 3816 & $3769.9\pm 2.5^{\,\dagger }$ & 0.07 & 0.97 \\ 
$3s$ & 4039 & $4040\pm 10^{\,\dagger }$ & 0.30 & 1.41 \\ 
$2d$ & 4119 &  & 0.15 & 1.47 \\ 
$4s$ & 4286 & $4159\pm 20^{\,\dagger }$ & 0.30 & 1.92 \\ 
$3d$ & 4341 &  & 0.18 & 2.00 \\ 
$5s$ & 4467 & $4415\pm 6^{\,\dagger }$ & 0.28 & 2.50 \\ \hline\hline
$1p$ & 3542 & $3525.3\pm 0.2$ & 0.11 & 0.72 \\ 
\end{tabular}
\end{table}

\begin{table}[tbh]
\caption{Leptonic widths $\Gamma _{e^{+}e^{-}}$ (in keV) for $c\overline{c}$
with the same model as in Table \ref{t8}. We denote by a '$\dagger $' the
experimental states with different possibilities for the orbital angular
momentum, it could be $L=0$ and/or $L=2$. Experimental data are taken from
Ref. \protect\cite{PDG}.}
\label{t9}
\begin{tabular}{ccccc}
& $\Gamma _{e^{+}e^{-}}^{(0)}\left( 1-{\frac{{16\alpha _{s}}}{{3\pi }}}%
\right) $ & $\left( \Gamma _{e^{+}e^{-}}\right) _{exp}$ & $\Gamma
_{e^{+}e^{-}}^{(0)}/\Gamma _{e^{+}e^{-}}^{(0)}(1s)$ & $\left[ \Gamma
_{e^{+}e^{-}}/\Gamma _{e^{+}e^{-}}(1s)\right] _{exp}$ \\ \hline
$1s$ & 2.30 & $5.3\pm 0.4$ & 1 & 1 \\ 
$2s$ & 0.87 & $2.19\pm 0.15$ & 0.38 & $0.41\pm 0.06$ \\ 
$1d$ & $1.3\times 10^{-2}$ & $0.26\pm 0.04^{\,\dagger }$ & $5.5\times
10^{-3} $ & $(5.0\pm 1.1)\times 10^{-2^{\,\dagger }}$ \\ 
$3s$ & 0.47 & $0.75\pm 0.15^{\,\dagger }$ & 0.20 & $0.15\pm 0.04^{\,\dagger
} $ \\ 
$2d$ & $1.72\times 10^{-2}$ &  & $7.4\times 10^{-3}$ &  \\ 
$4s$ & 0.34 & $0.77\pm 0.23^{\,\dagger }$ & 0.15 & $0.15\pm 0.05^{\,\dagger
} $ \\ 
$3d$ & $1.74\times 10^{-2}$ &  & $8.2\times 10^{-3}$ &  \\ 
$5s$ & 0.24 & $0.47\pm 0.10^{\,\dagger }$ & 0.10 & $0.09\pm 0.03^{\,\dagger
} $%
\end{tabular}
\end{table}

\end{document}